# Transfer Learning in Vocal Education: Technical Evaluation of Limited Samples Describing Mezzo-soprano

Zhenyi Hou[1]*[ƒ]   Xu Zhao[1]*   Kejie Ye[1]   Xinyu Sheng[1]   Shanggerile Jiang[1]   Jiajing Xia[1]

Yitao Zhang[1]   Chenxi Ban[1]   Daijun Luo[1]   Jiaxing Chen[1]   Yan Zou[3]

Yuchao Feng[2][ƒ]   Guangyu Fan[1]   Xin Yuan[2][ƒ]

[1]University of Shanghai for Science and Technology, Shanghai, 200093, China

[2]Westlake University, Hangzhou, 310024, China

[3]Shanghai Conservatory of Music, 200031, China

*These authors contributed equally

[ƒ]Corresponding author. E-mail:{batohou@hotmail.com; fengyuchao@wioe.westlake.edu.cn; xyuan@westlake.edu.cn}

## Abstract

Vocal education in the music field is difficult to quantify due to the individual differences in singers' voices and the different quantitative criteria of singing techniques. Deep learning has great potential to be applied in music education due to its efficiency to handle complex data and perform quantitative analysis. However, accurate evaluations with limited samples over rare vocal types, such as Mezzo-soprano, requires extensive well-annotated data support using deep learning models. In order to attain the objective, we perform transfer learning by employing deep learning models pre-trained on the ImageNet and Urbansound8k datasets for the improvement on the precision of vocal technique evaluation. Furthermore, we tackle the problem of the lack of samples by constructing a dedicated dataset, the Mezzo-soprano Vocal Set (MVS), for vocal technique assessment. Our experimental results indicate that transfer learning increases the overall accuracy (OAcc) of all models by an average of 8.3%, with the highest accuracy at 94.2%. We not only provide a novel approach to evaluating Mezzo-soprano vocal techniques but also introduce a new quantitative assessment method for music education.

## 1. Introduction

In vocal education, quantitative assessment of vocal techniques has always been an issue that remains unsolved. This issue mainly results in the diversity of educators with their teaching methods[1] and learners' differences in voice and physical conditions[2]. These factors hinder the development of systematic quantitative assessment, especially in evaluating rare voice types such as Mezzo-soprano, Countertenor, and Bass[3].

From the physiological viewpoint, voice types are determined by several factors, including the length, thickness, and tension of the vocal folds, as well as the size and configuration of the oropharyngeal cavity. For example, mezzo-sopranos usually have slightly longer and thicker vocal cords, enabling them to produce a lower voice. Because of this, they own a wide range, deep timbre, and excellent performance in the lower-middle register. Few female singers have this unique kind of vocal condition. In addition, due to the particular structure of the vocal folds, the mezzo-sopranos(compared to the sopranos) require more support from the diaphragmatic and the core respiratory muscles in order to maintain a sustained vibration of the vocal folds at specific pitches[4–6]. The difficulties of technical training further exacerbate the scarcity of mezzo-sopranos. The Mezzo-soprano has a high bias in chest voice (authentic voice), which makes the normalization of high and low registers and training over the voice change region more complex than other types.

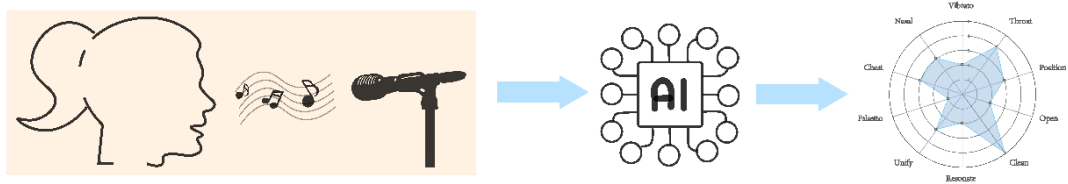

**Figure 1**. An illustration of vocal techniques evaluation. We record music clips of professional bel canto singers for training deep learning models, and then we apply them to evaluate the technical scores of unlabeled music.

In traditional vocal assessment, teaching and evaluation over-reliance on the educator's experience and subjective judgment, making it difficult to objectively and accurately reflect the real level of the learner's singing ability[7]. In this context, deep learning techniques demonstrate significant potential for application, especially in processing audio data and quantitative analysis of metrics[8,9]. The process is shown in **Fig. 1**.

In recent years, there has been tremendous progress in applying deep learning to audio classification[10–15]. Convolutional neural networks (CNNs) have demonstrated strong feature extraction capabilities in audio classification[10]. The convolutional recurrent network (CRNN) proposed by Choi et al. significantly improved classification accuracy[11]. Zhang et al. demonstrated that CNNs have achieved comparable accuracy to human beings in music genre classification achieved accuracy comparable to humans[12], and Stefanus et al. achieved accurate classification of choir vocal music by CRNN[13], further extending the application of deep learning to audio tasks. Demir et al. applied CNN on an environmental sound dataset and achieved good classification results[14]. Guzhov et al. proposed the ESResNet network[15] to improve environmental sound classification performance further. All of the above works extend the application of deep learning on audio data. However, they all rely on many well-labeled datasets and still have limitations when dealing with scarce vocal data with few samples.

Existing publicly available audio datasets are designed to cover a broad range of vocal types but remain under-represented for specific Bel canto voice parts, such as mezzo-sopranos[14,16–20]. Large-scale audio datasets like VGGSound and FSD50K provide valuable resources for audio classification studies[18,20], but they lack sufficient data to represent rare vocal types in the field of Bel canto. Existing professional vocal datasets are also inconsistent in labeling important points such as pitch, timbre, and technique, making them inefficient to use for quantitative analysis[21].

The lack of sufficient data has posed a significant challenge to apply deep learning in audio classification[20]. Transfer learning emerges as an effective solution [22–25]. Transfer learning is a method of extending a training model for a specific task to another target task to extract valid features for the new task based on the prior knowledge of the source task[25,26]. In recent years, transfer learning has been widely applied in small-sample learning[24,25,27], image processing[28–31] and audio classification[22,23,27].

To address the above issues, we constructed the Mezzo-soprano Vocal Set (MVS) dataset firstly. Then, we used models pre-trained on publicly available datasets, such as UrbanSound8K and ImageNet, to initialize the weights of our deep learning model[32–34]. Finally, we transferred the model to the MVS dataset for fine-tuning, which enables efficient learning of mezzo-soprano vocal technique features. Our approach not only effectively mitigates the issue of data scarcity but also significantly improves the accuracy of vocal technique evaluation. The main process of our method is shown in **Fig. 2**.

In summary, the main contributions of our work are as follows:

1. First, we construct a mezzo-soprano dataset for the audio evaluation of scarce vocal parts and carefully label ten professional vocal techniques according to a unified evaluation criterion.
2. We select three CNN-based deep learning models for audio evaluation[35–37] and further improve their generalization performances through transfer learning of pre-trained models for large datasets. Our experimental results show that our method improves the overall accuracy (OAcc) of all models by an average of 8.3%, with a maximum accuracy of 94.2%.
3. Our work provides an effective solution for the technical evaluation of the mezzo-soprano voice type and a personalized and efficient teaching tool for voice educators. At the same time, learners can engage in self-reflection and active learning according to the objective scores predicted by the model and develop independent learning techniques.

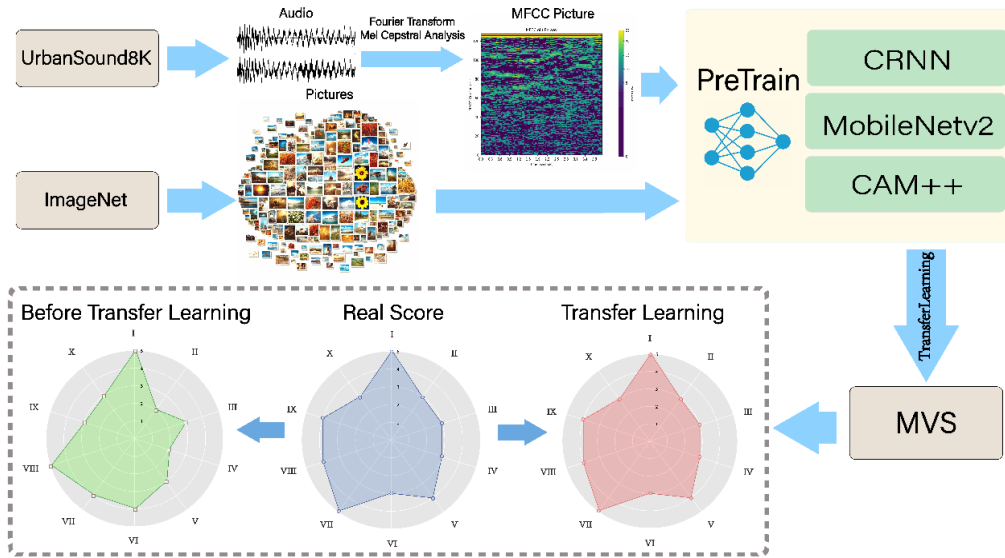

**Figure 2**. An illustration of the process of our method. We first pre-train CRNN, MobileNet v2, and CAM++ models on ImageNet and Urbansound8K datasets, and then transfer learning to our Mezzo-soprano Vocal Set (MVS) for fine-tuning to improve the accuracy of technical evaluation.

## 2. Proposed method

2.1 Rare voice type dataset

In order to study the assessment of vocal technique for rare voices[38], we construct the Mezzo-soprano Vocal Set (MVS) dataset focusing on Mezzo-soprano, which collects recordings of several professional singers in vocal singing, as well as singers majoring in vocal performance at the Conservatory of Music and obtains 1,212 high-quality audio segments of mezzo-sopranos. Each segment lasts for three to five minutes, sampled at 48,000 Hz, and saved as a WAV file. MVS aims to comprehensively characterize and evaluate the critical technical features of mezzo-soprano singing, especially in vocal training and education. The audio segments in MVS are labeled under 10 vocal techniques: Vibrato, Throat, Position, Open, Clean, Resonate, Unify, Falsetto, Chest, and Nasal. Each technique is rated on a scale of 1 (highest) to 5 (lowest) based on how well the technique is accomplished. The brief description of each technique is listed as follows:

**Vibrato.** Vibrato in vocal music refers to the vibrato of the voice. A relaxed and rounded state of muscles and an unobstructed breath produce a standard, beautiful vibrato. Unscientific vocalization, such as muscle tension and tongue pressure, can result in a fast or slow vibrato or a dull, straight sound without vibrato.

**Throat.** The throat plays a key role in controlling the airflow, which means that a right position of the throat can help singers make full use of the airflow to produce enough energy for volume control and regulation. At the same time, a stable throat can help singers maintain the stability and durability of their voices, and keep the quality and expressiveness over a long period of time.

**Position.** In the field of Bel canto, position refers to the singer's adjustment of the position of the throat to control the flow of the air. Singers need to learn to put their voices into the right places, such as specific areas of the head, nose or mouth, to achieve optimal resonance and vocal clarity.

**Open.** Cavity opening refers to the state in which the singer's throat and respiratory system are adjusted to allow the voice to vibrate and resonate freely in the mouth and throat.

**Clean.** Clean in vocal music refers to the purity and clarity of the sound and how accurately it matches the pitch. This concept is closely related to the quality and accuracy of the sound.

**Resonate.** Resonance is the process of utilizing body cavities to enhance the timbre and volume of a sound. It includes the resonance of the oral, pharyngeal, nasal, and head and chest cavities. Right throat position and opening can help a singer achieve pharyngeal resonance, resulting in a fuller and more powerful voice.

**Unify.** Experts pay attention to the unity of the sound in each voice area in the field of Bel canto. The principle of unity of the voice area of Bel canto requires the consistency in vocal method of the high, middle, and low voice areas, as well as the voice position and the timbre and volume of the voice.

**Falsetto.** Falsetto is the sound effect of the vibration of the edges of the vocal cords in singing. In the field of Bel canto, falsetto usually refers to those areas of the voice which are higher than the average chest voice, with a lighter, thinner sound texture, thus giving the audience an ethereal, floating feeling.

**Chest.** Chest in vocal music refers to the sound effect of the overall vibration of the vocal cords in singing. It is also known as full voice or chest voice and is a singing technique based on chest resonance. In the field of Bel canto, control of the chest voice is very important because it affects the expressiveness of the musical work and the singer's vocal health.

**Nasal.** The nasal tone in the singing timbre is wrong, resulting in muffled, unclear singing timbre, poor resonance, poor penetration, and other problems. We often apply binary standard to classify the nasal tone timbre aiming at judging the singer's singing quality.

To highlight the differences between MVS and the non-specialized dataset Urbansound8k, we transformed the audio data into MFCC spectrograms and compared them after visualization, as shown in **Fig. 3**.

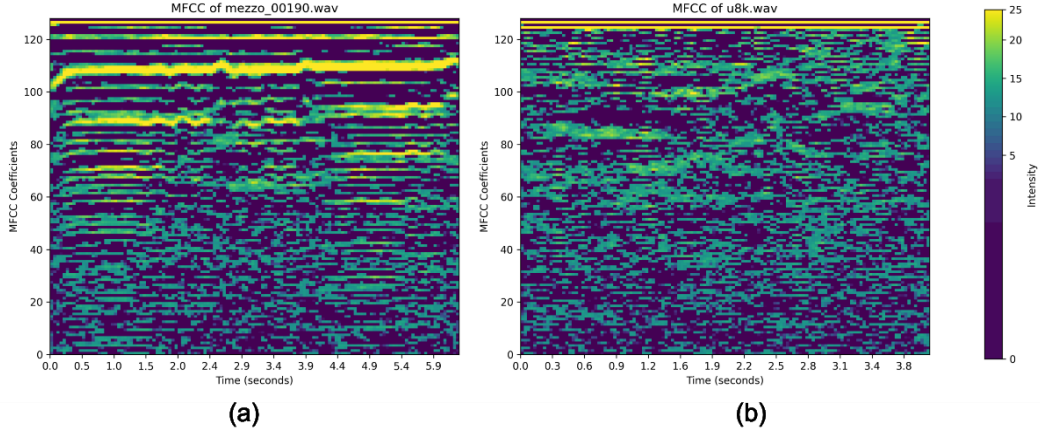

**Figure 3**. An illustration of MFCC spectrograms. (a) is a Mezzo-soprano spectrogram from MVS dataset, and (b) is a city sound from Urbansound8K dataset.

**Background noise**. The audio data in UrbanSound8K contains a lot of background noise, which is discrete and irregular. In contrast, it is entirely different in MVS, which intactly maintains the ripples of the human voice frequency and has less noise. The pure MVS dataset facilitates the model to focus on learning vocal frequency features. It is worth mentioning that this result does not mean that pre-training model on the UrbanSound8K dataset does nothing to improve the generalization performance of the model. Interestingly, the opposite is true. We will discuss our findings in the experimental section.

**Diversity of audio content**. The UrbanSound8K dataset contains different sounds in urban environments, such as car horns and barking dogs, which display large randomness and diversity in the spectrum. In contrast, Our MVS dataset consists of professional singing clips with more structured audio content, and the spectrum reveals a more stable harmonic structure and better pitch variation. The diversity in content means that although UrbanSound8K provides a rich set of basic audio features well suited for transfer learning, further fine-tuning is still essential to adapt to specific vocal features.

**Specialization of Labeling**. The UrbanSound8K dataset is extensively used for category differentiation of sound sources, with labels focusing on only ten sound categories. In contrast, our MVS dataset scores and labels ten specialized vocal techniques. Because of this, our labels are more refined. Our dataset improves the model's efficiency in learning higher-level and detailed musical features. Our result shows that specialized vocal datasets are significant for vocal technique assessment.

2.2 Mel-frequency Cepstral Coefficient

We preprocess the given audio file into Mel-frequency Cepstral Coefficient (MFCC) features[39] before further investigations. MFCC features are widely used in issues such as speech recognition, sentiment analysis, and voiceprint recognition. MFCC captures features containing sound information by simulating the working mode of the human ear, and its transformation process is defined by the following formula[40]:

$$mel(f) = 2595 * log\left[log_{10}\left(1 + \frac{f}{700}\right)\right] \quad (1)$$

where $f$ is a linear frequency. We first perform pre-emphasis filtering on the speech signal to enhance the energy of the high-frequency part and compensate for the high-frequency attenuation caused by the channel effect. Calculation is conducted according to the formula:

$$S'(n) = S(n) - \alpha * S(n-1) \tag{2}$$

where $\alpha$ is called the pre-emphasis coefficient, with a value of $0.9 \leq \alpha \leq 1$. After the calculation, we perform Hamming Window processing on each frame to minimize signal discontinuity at the beginning and the end of each frame caused by frame blocking process. As shown in the following formula:

$$w(n) = 0.54 - 0.46 \cos\left(\frac{2\pi n}{N-1}\right) \tag{3}$$

where $0 \leq n \leq N - 1$. Then we divide the speech signal into multiple short time frames for fast Fourier transform (FFT), converting the audio data into multiple sine wave function data to obtain the amplitude spectrum of each frequency component, as defined by the formula:

$$X_a(k) = \sum_{n=0}^{N-1} x(n) e^{-j2\pi k/N}, 0 \leq k \leq N \tag{4}$$

After that, a set of triangular filters is used to weight and average the amplitude spectrum to obtain the Mel frequency spectrum. These filters are evenly distributed on the Mel frequency scale, with denser filters in the low frequency range and sparser filters in the high frequency range:

$$H_m(k) = \begin{cases} 0, & k < f(m-1) \\ \frac{2(k-f(m-1))}{(f(m+1)-f(m-1))(f(m)-f(m-1))}, & f(m-1) \leq k \leq f(m) \\ \frac{2(f(m+1)-k)}{(f(m+1)-f(m-1))(f(m)-f(m-1))}, & f(m) \leq k \leq f(m+1) \\ 0, & 0 \end{cases} \tag{5}$$

$$\sum_{m=0}^{M-1} H_m(k) = 1 \tag{6}$$

Then, we take the logarithm of the output of the filter bank to obtain the logarithmic Mel frequency spectrum. This step imitates the perceptual characteristics of the human ear, due to the fact that the perception of sound intensity by the human ear is logarithmic:

$$s(m) = \ln\left(\sum_{k=0}^{N-1} |X_a(k)|^2 H_m(k)\right), 0 \leq m \leq M \tag{7}$$

Finally, the logarithmic Mel spectrum is subjected to discrete cosine transform (DCT) to convert it into cepstral coefficients, resulting in a set of MFCC coefficients. The result of this conversion can be calculated using a formula:

$$c(n) = \sum_{k=1}^{K} \log S_k * \cos\left(n\left(k - \frac{1}{2}\right)\frac{\pi}{K}\right) \tag{8}$$

where $n = \{1, 2, \cdots, K\}$, $S_k$ is the output of the filter bank at index $k$, and $K$ is the expected coefficient.

2.3 Transfer learning

We pre-trained the model on a large-scale dataset, and after multiple rounds of training, we have found a better local optimal solution from the loss function[26,41], making the loss function reach a minimum value:

$$\theta^* = argmin_\theta[\mathcal{L}(\theta)] \tag{9}$$

where the loss function of the pre-trained model is $\mathcal{L}(\theta)$. $\theta$ is the weight parameter of the model. During the fine-tuning process on the MVS dataset, we did not start with random initialization. Instead, we started training from the pre-trained model's $\theta^*$, and the new objective loss function is:

$$L_{new}(\theta^*) = \sum_{i=1}^{n} l(f(x_i, \theta^*), y_i) \tag{10}$$

where $L_{new}(\theta^*)$ is the new objective loss function, $f(x_i, \theta^*)$ is the model's prediction on the input data $x_i$, $l(\cdot)$ is the loss of a single sample, and $y_i$ is the target label. $\theta^*$ is the weight parameter obtained from the pre-trained model. At this point, the model can find out new local optima more easily with less training starting from $\theta^*$.

We pre-trained the model on a larger dataset to learn universal features. After we fine tune on smaller datasets, the model tends to find an optimal solution that balances generalization ability, rather than overfitting the details of a small number of samples. We use transfer learning methods to effectively transfer extensive knowledge learned on larger datasets to smaller datasets. This not only improves the performance of the model, but also effectively prevents overfitting and exhibits faster convergence speed during the optimization process.

**3. Experimental Setup**

In this section, we perform deep learning methods on supervised vocal technique assessment through the MVS dataset. First, in Section 3.1, we propose the problem formulation. Then, in Section 3.2, we briefly introduce the baseline model. Finally, in Sections 3.3 and 3.4, we elaborate on the experimental setup as well as the evaluation criteria for training and testing.

3.1 Problem definition

We assume a set of $n$ training vocal segments $V = \{V_i\}_{i=1}^n$ with corresponding labels $Y = \{Y_i\}_{i=1}^n$. $Y_i$ is a set of labels for the vocal techniques, where $n = 10$. We study the MFCC spectral characteristics of Bel canto singing to achieve scoring and appreciation of bel canto singing techniques.

3.2 Baseline

To provide a baseline model for comparison, we extracted MFCC features from audio files of the MVS dataset and trained three CNN models (CRNN[35], MobileNet v2[36], and CAM++[37]) without using data augmentation strategies to benchmark their performance.

3.3 Pre-trained Models

We use the same architecture, hyperparameter settings, and training program as it had been used in the baseline model. Three CNN-based networks, CRNN, MobileNet v2, and CAM++, were pre-trained on ImageNet and Ubransound 8k respectively. During the testing process, the predicted results of the dataset refer to the average of the accuracies from three separate training sessions of these three models.

The ImageNet dataset contains images from hundreds of object categories, totaling over 14 million images. It contains images of diverse objects and scenes, ranging from images of animals and plants to images of everyday objects. All images are accurately labeled, and each image has a corresponding object category and position label. This allows the dataset to provide accurate reference standards for training and evaluating algorithms.

The Urbansound8k dataset is a classic audio dataset containing 8732 detailed annotated urban sound clips, covering 27 hours of audio recording, all stored in WAV format at a sampling rate of 44100 Hz. It is often used to test the performance of sound classification and recognition algorithms, especially in complex urban environments. This dataset covers 10 different categories of urban sounds, such as air-conditioning, car horn, and children playing.

3.4 Training and evaluation

All of our models are implemented and trained on PyTorch[42]. We use Adam as the optimizer, with an initial learning rate of 0.0001, and train our model using the cross-entropy loss function.

During the training process, the batch size is 64. In addition, when OAcc(Overall Accuracy) did not improve after more than 10 verifications, we adopted an early stopping strategy to halt the training process. Finally, we evaluated it using the model that achieved the highest accuracy on the validation set. The random seeds for PyTorch and scikit learn are fixed to avoid biased results. We standardize the input data and convert it to a size of 224 × 224 to facilitate subsequent transfer learning. Finally, the MVS dataset was divided into training, validation, and testing subsets in an 8:1:1 ratio.

3.5 Metric

We use the highest accuracy on the validation set to evaluate, in which the formula for accuracy is defined as follows:

$$OAcc = \frac{TP}{TP+FP} \quad (11)$$

where $TP$ (True Positive) is the number of correctly predicted samples, and $FP$ (False Positive) is the number of incorrectly predicted samples.

## 4. Results

**Table 1**. Accuracy of deep learning models in audio evaluation on MVS dataset.

| Models | Baseline（OAcc） | ImageNet（OAcc） | Urbansound（OAcc） | Urbansound+ImageNet（OAcc） |
|---|---|---|---|---|
| CRNN | 80.5% | 81.2%（+0.7%） | 85.4%（+4.9%） | **87.3%（+6.8%）** |
| MobileNet v2 | 85.2% | 86.4%（+0.9%） | 89.2%（+4.0%） | **94.2%（+9.0%）** |
| CAM++ | 76.4% | 78.3%（+1.9%） | 80.3%（+3.9%） | **85.4%（+9.0%）** |

**Tab. 1** illustrates the audio evaluation results of the three models. Thanks to the efficient structure of MobileNet v2, it outperforms CRNN and CAM++ in all scenes. Comparing the baseline without loading the weights of the pre-trained models, all our models are upgraded and improved after pre-training with additional datasets (ImageNet or Urbansound8k). Among them, the increase in accuracy of the models after pre-training on the large-scale image dataset ImageNet is relatively small (0.7% to 1.9%). Meanwhile, pre-training on the Urbansound 8k audio dataset significantly improves the accuracy of the models, with an average increase of 4.3%. Specifically, CRNN shows an increase of 4.9%.

What is interesting is that when we combined Urbansound8k and ImageNet for pre-training, the performances of all models gain significantly. This is especially remarkable for MobileNet v2, which achieved an accuracy of 94.2%, for a 9.0% improvement over the benchmark. Besides, CAM++ gained 9.0%. In conclusion, all models showed a considerable performance improvement after transfer learning the pre-training weights of the additional dataset. It demonstrates the simplicity and effectiveness of the transfer learning for audio evaluation on the MVS dataset.

**Table 2**. Prediction accuracy of scores for ten vocal techniques by deep learning models on the MVS dataset.

| Models | Average Precision | Vibrato | Throat | Position | Open | Clean | Resonate | Unify | Falsetto | Chest | Nasal |
|---|---|---|---|---|---|---|---|---|---|---|---|
| CRNN | 87.3% | 86.5% | 87.2% | 87.7% | 88.3% | 86.8% | 86.7% | 87.4% | 88.1% | 87.1% | 87.3% |
| MobileNet v2 | 94.2% | 93.7% | 94.3% | 95.5% | 93.3% | 94.7% | 94.4% | 93.4% | 95.1% | 93.5% | 94.2% |

| | | | | | | | | | | |
|---|---|---|---|---|---|---|---|---|---|---|
| CAM++ | 85.4% | 85.8% | 85.9% | 86.1% | 84.7% | 85.5% | 86.8% | 83.9% | 84.9% | 85.1% | 85.3% |

The scoring accuracies of the three deep learning models for the ten categories are shown in **Tab. 2**. We take the average accuracy of all technique categories as the OAcc of the prediction results. From the table, we can intuitively see that transfer learning is crucial for deep learning models to accurately predict the mezzo-soprano singing technique scores.

To qualitatively analyze the changes brought by transfer learning to the model training and testing performance, we choose the most effective model, MobileNet v2, as an example. **Fig. 4** contains four subgraphs, each showing the changes in the performance metrics of MobileNet v2 during training and testing before and after transfer learning. Specifically, the subgraph in the first row shows how the model's accuracy and loss values change with epoch (number of training rounds) during training, while the second row shows the testing process.

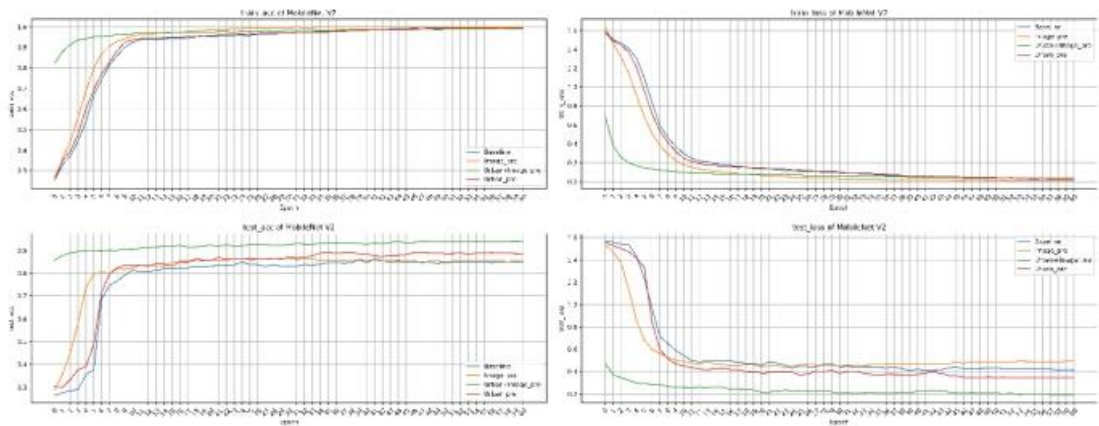

**Figure 4**. Loss and accuracy curve during training and testing process.

We zoom in on the Test Accuracy vs. Test Loss curves during testing (as shown in **Fig. 5**) to analyze the changes in model performance caused by transfer learning. Figure A shows the graph of test loss curve, with the blue curve as the baseline. As highlighted by the red frames in **Fig. 5a** and **5b**, the model converges significantly faster after learning the pre-training weights of ImageNet[43], as shown by the orange curve.

However, after transfer learning the pre-training weights of Urbansound8k (red curve in **Fig. 5a**), the model's convergence speed did not change significantly, even though its loss decreased. We have found that training the model on datasets with background noise is beneficial to improve robustness[17,25,26]. Since adding noise is a standard way for data enhancement, previous works[19,28,30] support this idea. Thus, to some extent, learning the Urbansound 8k pre-training weights improves the performance of the model.

After we merge the two pre-training weights for transfer learning, the convergence speed improves by leaps and bounds, as shown by the green curve in the blue frame in **Fig. 5a**, which is smooth and with minimal loss. As the number of training increases, the orange curve based on ImageNet pre-training gradually becomes overfitted, as shown in the blue frame in **Fig. 5a**. However, the generalization performance of the red curve based on Urbansound8k is stable. We believe that the pre-training weights of ImageNet help the model to "warm up" quickly to adapt to the training, but do not improve the generalization performance of the model. Therefore, during redundant training, performance gradually degrades, and overfitting occurs.

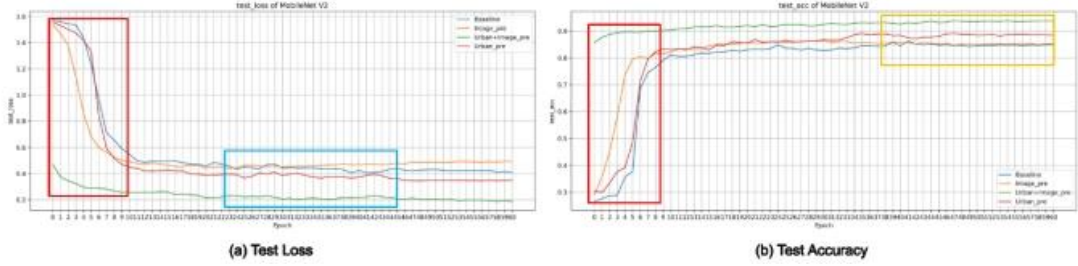

**Figure 5**. Loss and accuracy curve during testing process. (a) is the test loss curve, and (b) is the test accuracy curve.

For the test accuracy, as shown in **Fig. 5b** with the orange frame highlighted, the model slightly improved its accuracy after transfer learning through the audio dataset Urbansound8k. However, after combining the image dataset ImageNet, a vast accuracy improvement was obtained, as shown by the green curve. Moreover, the highest accuracy did not change through transfer learning on ImageNet, as shown by the orange curve. We demonstrated that the joint transfer learning of the pre-trained model weights of the image and audio datasets can significantly improve the generalization performance of the prediction model. However, it remains challenging to explain the feature changes in the model before and after transfer learning[44].

Therefore, we use Grad-CAM[45] to visualize the gradient-weighted class activation maps of the features in the deep learning model through qualitative analysis to explain the effect on the learning of MFCC features in the deep learning model before and after transfer learning.

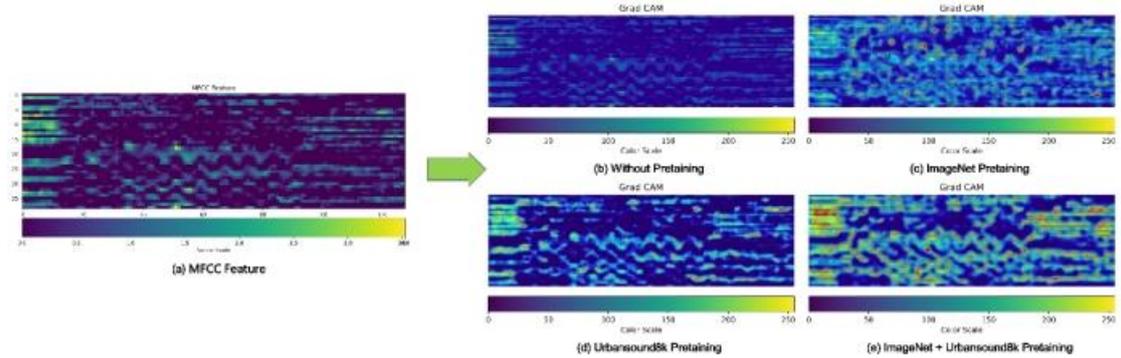

**Figure 6**. Visualization of Gradient-weighted Class Activation Mapping (Grad-CAM) for MFCC feature.

The first column in **Fig. 6a** shows the MFCC feature as input to the network, and the second and third columns show the gradient class activation map (Grad-CAM) of the input feature. The Grad-CAM visually tells us which regions the model focuses on during its process of learning features.

When no pre-training is performed, the model is insensitive to the regional distribution of the features, and no bright regions are visible in **Fig. 6b**. After introducing the pre-training weights for the ImageNet dataset, the model starts to focus on the regional distribution of the features, as shown by the bright regions in **Fig. 6c**, where the network focuses on the regions with higher energy distribution in the spectrogram. It attempts to learn the features in these regions, just as it learns to detect regions around objects in an image. Thus, the pre-trained ImageNet model can be extended to audio spectrograms. However, the bright regions are cluttered and cannot accurately focus on audio features (as shown in **Fig. 6d**). After we pre-trained the model using the Urbansound 8k dataset, the model obtained a clear field of view and detected audio features with higher energy as well as fewer red regions and lower sensitivity. Things began to change after we pre-trained the model jointly with ImageNet and Urbansound8k, the model was sensitive to the regions of each audio feature (regardless of the energy level) and learned the coarse and fine-grained feature information (as shown in **Fig. 6e**).

From these experiments, we can intuitively find out that transfer learning is essential for he process of learning audio data with a small number of samples. These results show that even by learning between two different domains of image and audio data, we can still improve the network's generalization performance. Moreover, the effect of joint transfer learning of the two is even more significant.

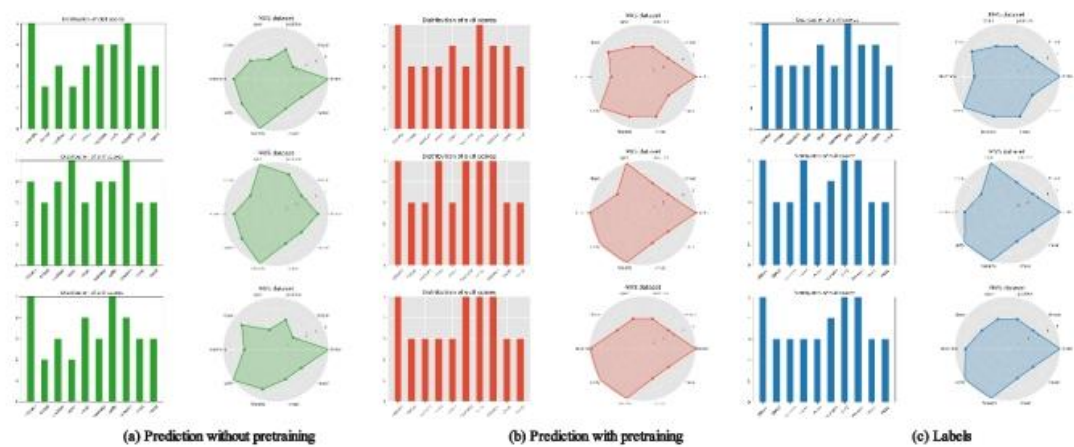

**Figure 7**. Comparison of model prediction results.

The visualization of the prediction results for three sets of audio evaluation is shown in **Fig. 7**. Each data set contains a histogram and a radar plot. These plots show the distribution of technique scores after different training schemes. The first column (**Fig. 7a**) shows the prediction results without pre-training, the second column (**Fig. 7b**) shows the prediction results of the model that underwent joint pre-training with ImageNet and Urbansound8k, and the last column (**Fig. 7c**) shows the labels.

Comparing the plots of the first column with those of the second column, we can see that the transfer learning resulted in changes in the scores for all categories. If we compare the results in the second column with the labels in the third column, we can see that the model predicts each category more accurately. It also demonstrates that transfer learning improves the generalization performance of the model.

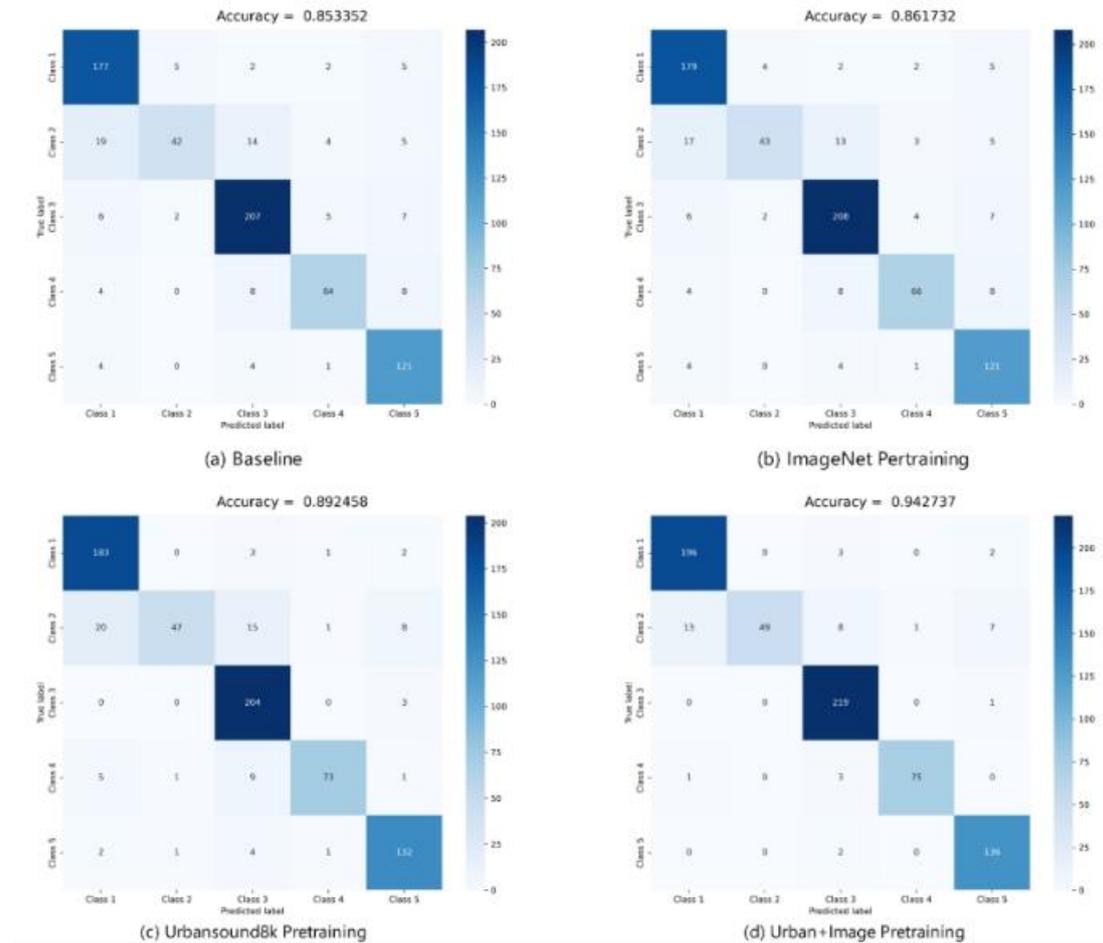

**Figure 8**. Comparison of confusion matrix.

Fig. 8 shows four confusion matrices corresponding to the predictions under different model training conditions. Each confusion matrix shows the performance of the model in the assessment, and the accuracy of the confusion matrices is shown at the top of each matrix. After transfer learning, the model's prediction accuracy gradually increases. In particular, the combination of Urbansound8k and ImageNet resulted in a model prediction accuracy of up to 94.2%. The confusion between 2-point- and 3-point results is most evident in the baseline.

In contrast, it is gradually reduced after pre-training. By analyzing these four confusion matrices, we can conclude that pre-training can significantly improve the scoring performance of deep learning models. The combined pre-training strategy (using ImageNet and Urbansound 8k dataset) improves accuracy, making the model perform even better in various conditions.

## 5. Discussion

1. Impact of transfer learning image datasets on models

We consider the ImageNet pre-trained model to be an excellent texture detector that can be effectively extended to MFCC spectrograms, as can be seen from Grad-CAM (**Fig. 6**), which is sensitive to the energy waviness of the audio and adaptively detects the texture of the audio. It helps the model to quickly capture audio features, as seen in the test curve (**Fig. 5**).

2. Impact of the transfer learning audio dataset on models

The variety of environmental sounds in the UrbanSound8K dataset and the vocal data in our MVS makes it easier for the model to explore the common features of the sounds, even if they are differences in content. The background noise improves the robustness of the model to some extent as well.

3. Impact of joint transfer learning on models

We believe that joint transfer learning is not a simple "1+1=2" process, but rather a multimodal feature adaptation process. The model learns different dimensions of audio features. We will conduct extensive experiments and deeply explore the principles and applications in the field of multimodality in our future investigations.

4. Future works and improvements

In the future, we first aim to increase the size and volume of the dataset for robust training and exceptionally more extensive vocal material. Then, we will further improve the quality of the labels and refine the score distribution. We will continue to keep pace with advanced transfer learning methods and use more robust models to improve the quality of audio scoring in further investigations. At the same time, we would like to further expand the application scenarios of singing technique evaluation by adding musical emotions to the evaluation system, since music is rich in multidimensional information, and the Bel canto is an artistic crystallization combining music and emotion.

## 6. Conclusion

In this paper, we construct a Mezzo-soprano Vocal Set (MVS) dataset to quantify and evaluate vocal techniques by deep learning. At the same time, we pre-train models on ImageNet and Urbansound8k datasets to tackle the issues of low prediction accuracy and poor generalization performance of the evaluation models due to the scarcity of data samples. Our experimental results show that after transfer learning, the OAcc of all models improved by 8.3% on average, with the highest model reaching 94.2%. For the Mezzo-soprano, which is a rare voice type, our work not only provides a scientific method for vocal technique assessment, but also provides educators with a more efficient tool for personalized and effective vocal teaching.

**Data Availability Statement**

The publicly available datasets used for transfer learning in this study can be found at https://urbansounddataset.weebly.com/urbansound8k.html and https://www.image-net.org/. Our self-constructed mezzo-soprano dataset, MVS, can be obtained from the corresponding author upon reasonable request.